\documentclass[12pt]{spieman}  
\usepackage{amsmath,amsfonts,amssymb}
\usepackage{graphicx}
\usepackage{float}
\usepackage{setspace}
\usepackage{tocloft}
\usepackage{gensymb}

\title{HODET: Hybrid Object DEtection and Tracking using mmWave Radar and Visual Sensors}

\author[a]{Joseph St. Cyr}
\author[b]{Joshua Vanderpool}
\author[a,*]{Yu Chen}
\author[a]{Xiaohua Li}
\affil[a]{Dept. of Electrical \& Computer Engineering, Binghamton University, Binghamton, NY 13902}
\affil[b]{The Raymond Corporation, Greene, NY 13778}

\cftpagenumbersoff{figure}
\cftpagenumbersoff{table} 

\begin{document} 
\maketitle

\begin{abstract}
Image sensors have been explored heavily in automotive applications for collision avoidance and varying levels of autonomy. It requires a degree of brightness, therefore, the use of an image sensor in nighttime operation or dark conditions can be problematic along with challenging weather such as fog. Radar sensors have been employed to help cover the various environmental challenges with visible spectrum cameras. Edge computing technology has the potential to address a number of issues such as real-time processing requirements, off-loading of processing from congested servers, and size, weight, power, and cost (SWaP-C) constraints. This paper proposes a novel Hybrid Object DEtection and Tracking (HODET) using mmWave Radar and Visual Sensors at the edge. The HODET is a computing application of low SWaP-C electronics performing object detection, tracking and identification algorithms with the simultaneous use of image and radar sensors. While the machine vision camera alone could estimate the distance of an object, the radar sensor will provide an accurate distance and vector of movement. This additional data accuracy can be leveraged to further discriminate a detected object to protect against spoofing attacks. A real-world smart community public safety monitoring scenario is selected to verify the effectiveness of HODET, which detects, tracks objects of interests and identify suspicious activities. The experimental results demonstrate the feasibility of the approach.

\end{abstract}

\keywords{Hybrid Detection and Tracking, mmWave Radar, Visual Sensor,  Convolutional Neural Network.}

{\noindent \footnotesize\textbf{*}Corresponding Author: Yu Chen,  \linkable{ychen@binghamton.edu} }


\section{Introduction}

The unprecedented pace of urbanization poses many opportunities and challenges \cite{chen2016dynamic}. The recent concept of Smart Cities has attracted the attention of the urban planners and researchers to enhance the security and well-being of the residents \cite{chen2018smart, chen2017smart}. One of the most essential smart community services is the intelligent resident surveillance \cite{nikouei2018smart, wu2017container}. It enables a broad spectrum of promising applications, including access control in areas of interest, human identity or behavior recognition, detection of anomalous behaviors, interactive surveillance using multiple cameras and crowd flux statistics and congestion analysis and so on \cite{hu2004survey}. 

There is a considerable amount of interesting research being conducted in the field of real-time object identification and tracking \cite{xu2019blendmas, xu2018real}.  In order to identify and track an object in real-time a few requirements need to be defined. One obvious requirement is that there must be some sort of interesting object (or objects) that is worth identifying and tracking. Another is that a system must be defined that will track and identify such an object \cite{nikouei2019kerman, wu2014container}. A typical component of one of these systems is an image sensor. The image sensor allows the system to ``see'' the object.  Once the object can be seen, the system needs to be able to process the image it is ``seeing'' in order to identify, track and possibly make other decisions. A common term to describe this technology is machine vision.

Image sensors have been explored heavily in automotive applications for collision avoidance and varying levels of autonomy \cite{hosticka2003cmos}. Complementary metal-oxide-semiconductor (CMOS) imagers convert photons to a proportional voltage that is read by the sensor to digitise the scene. This process requires a degree of brightness therefore use of an image sensor in nighttime operation or dark conditions can be problematic along with challenging weather such as fog \cite{bigas2006review}. Radar sensors have also been employed to help cover the various environmental challenges with visible spectrum cameras \cite{gessner2015multi}. The frequency range of the radar sensor is an important characteristic when selecting an appropriate sensor for an application.  For instance, Earth’s atmosphere contains a considerable amount of water vapor which leads to a high level of frequency absorption in the 60 GHz range. Typically, to overcome this limitation a higher frequency, such as 77 GHz, is utilized \cite{marcus2005millimeter}. An additional benefit of using a higher frequency is that it provides a finer range resolution.

An adjacent and popular topic of research is edge computing \cite{ahmed2017mobile, shi2016edge}. This technology has the potential to address a number of issues such as real-time processing requirements, off-loading of processing from congested servers, and size, weight, power and cost (SWaP-C) constraints \cite{allen2017strategies, nikouei2019decentralized, nagothu2018microservice} to name a few. Much of the edge computing phenomenon is being driven by the ever-growing Internet-of-Things (IoT) \cite{satyanarayanan2017emergence}. It is expected that the IoT trend will continue to dominate the future of the Internet and our data processing paradigms for many years to come \cite{wollschlaeger2017future}. 

An application of the previously mentioned image and radar sensor data is object detection and classification (e.g. provide detection of objects and count the number objects classified as a human that occupy, enter, or exit the field of view of the sensors). There are established methods and algorithms to facilitate the decision making a system must compute to detect and classify objects \cite{nikouei2019safe}. A few examples of the open source software libraries available are OpenCV \cite{bradski2008learning}, TensorFlow \cite{abadi2016tensorflow}, Saliency \cite{zhao2015saliency}, and YOLO \cite{redmon2016you}. These libraries make use of several techniques to detect and classify an object (e.g point and edge detection, machine learning, neural networks). An effective method to quickly detect and classify an object is to deploy machine learning with the use of a Convolution Neural Network (CNN) \cite{nikouei2018lcnn}.

This paper proposes a novel hybrid object detection and tracking (HODET) using two types of sensors, mmWave radar and visual sensor. The HODET is an edge computing application of low SWaP-C electronics performing object detection, tracking and identification algorithms with the simultaneous use of image and radar sensors. While the machine vision camera alone could estimate the distance of an object, the radar sensor will provide an accurate distance and vector of movement. This additional data accuracy can be leveraged to further discriminate a detected object to protect against spoofing attacks.

The rest of the paper is organized as follows: Section \ref{sec:related work} provides a review of related work. Section \ref{sec:algorithm} presents our proposal of the hybrid detection and tracking using radar and vision sensors and the technical approach. Experimental results are presented in Section \ref{sec:experimental}. Finally, Section \ref{sec:conclusions} wraps up this paper with the conclusions.


\section{Motivation and Related Work}
\label{sec:related work}

This section briefly discusses some of the related work that has been completed in closely related applications that inspired our proposed HODET system. Basically, the HODET is an automotive application that draws upon a collection of object detection, tracking and identification algorithms. It leverages the knowledge and insights under the umbrella of edge computing.

A sensor fusion based pedestrian collision detection system was proposed, which used a monocular CMOS camera as the image sensor along with a millimeter-wave radar sensor \cite{suzuki2010sensor}. Data from the radar sensor was used with a Kalman filter to provide detection and tracking of objects. The image sensor was used to detect whether or not a pedestrian crosswalk was present. Combining these two data streams allowed them to calculate the probability of a pedestrian collision and trigger a warning system based on a threshold.  

Another reported solution with pedestrian detection as its primary goal uses similar methods and highlights their use of a model-based detection algorithm \cite{milch2001pedestrian}. Meanwhile, in a research vehicles and and bicycles are added as primary objects to be detected in addition to pedestrian \cite{sugimoto2004obstacle}. This solution uses a charge-coupled device (CCD) camera with a millimeter-wave radar. They pointed out that, ``\emph{millimeter-wave radar offers advantages of higher reliability in bad weather conditions}'' in contrast to longer-range radars such as Light Detection and Ranging (LIDAR). 

The motivation of using multiple different types of sensors (i.e. image and radar) was summarized as ``[A] \emph{camera provides high spatial resolution but low accuracy in estimation of the distance to an object.  The high spatial resolution of the camera can support the low directional resolution of the radar, and the high distance resolution of the radar can support the low accuracy in distance estimation of the camera}.'' \cite{sugimoto2004obstacle} A deep convolutional neural network is proposed as a method of sensor fusion and object classification for autonomous vehicles \cite{gao2018object}.  

In contrast to the previously mentioned solutions, a system was proposed that use a red/green/blue (RGB) color camera data and two different types of radar sensors, millimeter-wave and LIDAR \cite{gao2018object}. The data from these sensors are fused and fed into an AlexNet \cite{alom2018history} to classify pedestrians, bicycles, cars and trucks. After a supervised training, the deep CNN method provided an efficient and accurate classification of their four primary objects of interest. Meanwhile, a color and thermal camera was used for image sensing in conjunction with millimeter-wave and LIDAR radar sensors to deploy a self-driving vehicle \cite{cho2014multi}. The thermal image sensor is used to ``\emph{perceive objects in challenging driving conditions, such as at night and in fog}.'' They propose detecting and classifying objects using motion and observation models of their primary objects of interest (i.e. pedestrians, bicyclists, vehicles).  

Another solution based on image and radar sensor fusion focused on pedestrian safety, which used sensor fusion at two levels (low and high) of their architecture \cite{tons2004radar}. A symmetrical deep convolutional neural network is used to detect changes in heterogeneous images taken at different times and dates by optical and radar sensors for the purpose of urban growth tracking, land use monitoring, and disaster evaluation \cite{liu2016deep}. Although it is not directly related to an automotive application like the other papers mentioned above, it is still applicable to our work. An interesting aspect of their work is the employment of unsupervised CNN training.

It would be remissive of us to not highlight one of the critical tasks we will need to address as part of our research: sensor calibration. Each of the sensors in our system will produce data in its own local coordinate system. To use the sensor data effectively we will need to calibrate the image and radar sensor data to exist in a common coordinate system. This is necessary to accurately detect, track and classify objects in the field of view (FOV). Sensor calibration is mentioned in all of the papers previously discussed \cite{suzuki2010sensor, milch2001pedestrian, sugimoto2004obstacle, gao2018object, cho2014multi, liu2016deep} except one \cite{tons2004radar}, which does not explicitly mention calibration however it is assumed. In general, a transformation matrix is used to translate data points from one coordinate system to another, although some slightly different methods are presented for obtaining the transformation matrix.  Primarily, the transformation matrix was obtained through some method of data point collection and correlation.  

\section{HODET: Hybrid Object DEtection and Tracking}
\label{sec:algorithm}

\subsection{System Architecture}

\begin{figure}[t]
   \centering
        \includegraphics[width=0.98\textwidth]{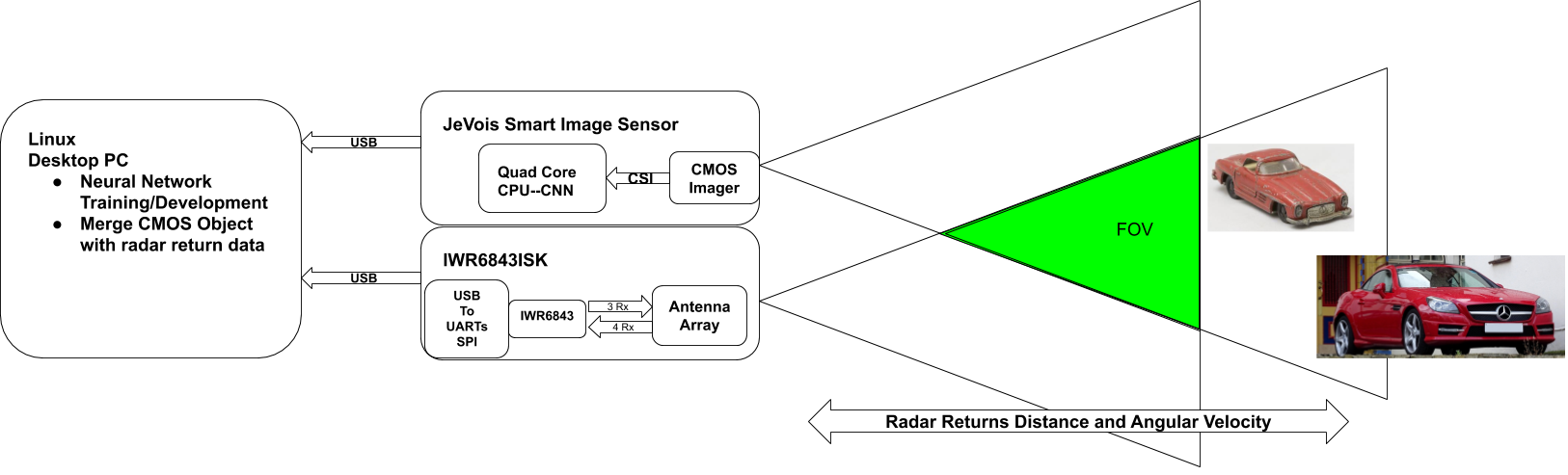}
   \caption{A System Level Illustration of HODET.}
    \label{fig:block}
\end{figure}

Figure \ref{fig:block} shows the system architecture of the proposed HODET scheme, which is a system constructed of a Linux PC to collect and merge data streams from the CMOS \emph{smart} image sensor and mmWave radar returns. The JeVois Smart Machine Vision Camera \cite{jevois2020jevois, jevois2020jevoisb} allows the use of the quad-core CPU to run one level of algorithms to identify objects. The radar sensor provides returns to help discriminate objects identified by the CMOS imager. The Radar could also be used in cases where it is not conducive for image sensors such as night time or in a climate weather. This system is built to explore the possibilities and would/could be merged into a more powerful SoC to run a centralized neural network at the edge. The diagram in Fig. \ref{fig:block} also illustrates the preliminary configuration where the JeVois and radar data will be collected, correlated, and processed by a development PC running the CNN algorithm.

Our technical approach will be incremental, building upon a baseline implementation.  The JeVois Smart Machine Vision Camera has been chosen as our SWaP-C friendly image sensor and processor. JeVois is an open source machine vision camera developed by Professor Laurent Itti and his team at the University of Southern California (USC) \cite{jevois2020jevoisc}. This camera has a substantial amount of user demo modules which can be used to familiarize ourselves with the product and establish a baseline implementation.  There is also an established support community \cite{jevois2020jevois} which will be useful for debugging help. JeVois also has beta development tools and an extensive code base stored in GitHub \cite{jevois2020jevoisd} which can be leveraged for rapid development.

Once the baseline is established the radar sensor will be integrated into the solution. Texas Instrument’s TIDEP-01000 millimeter-wave radar sensor \cite{texas2018people} has been selected. The radar sensor will provide accurate range, velocity, and angle information of the objects thus allowing further discrimination; Image sensors alone could be fooled with pictures or 3D models of objects. The JeVois will provide results via the USB webcam output and on the UART (serial port).  This data will be merged with the radar sensor point cloud or object field given in three dimensions (i.e. X,Y,Z-axis coordinates).  The radar’s field of view is 120\degree~horizontal and 30\degree~vertical.  As previously mentioned the views will not be collinear nor matched in the FOV. Some transformations will be needed and accounted for using a calibration methodology. 

\subsection{Calibration Method}

\begin{figure}[t]
    \centering
        \includegraphics[width=0.8\textwidth]{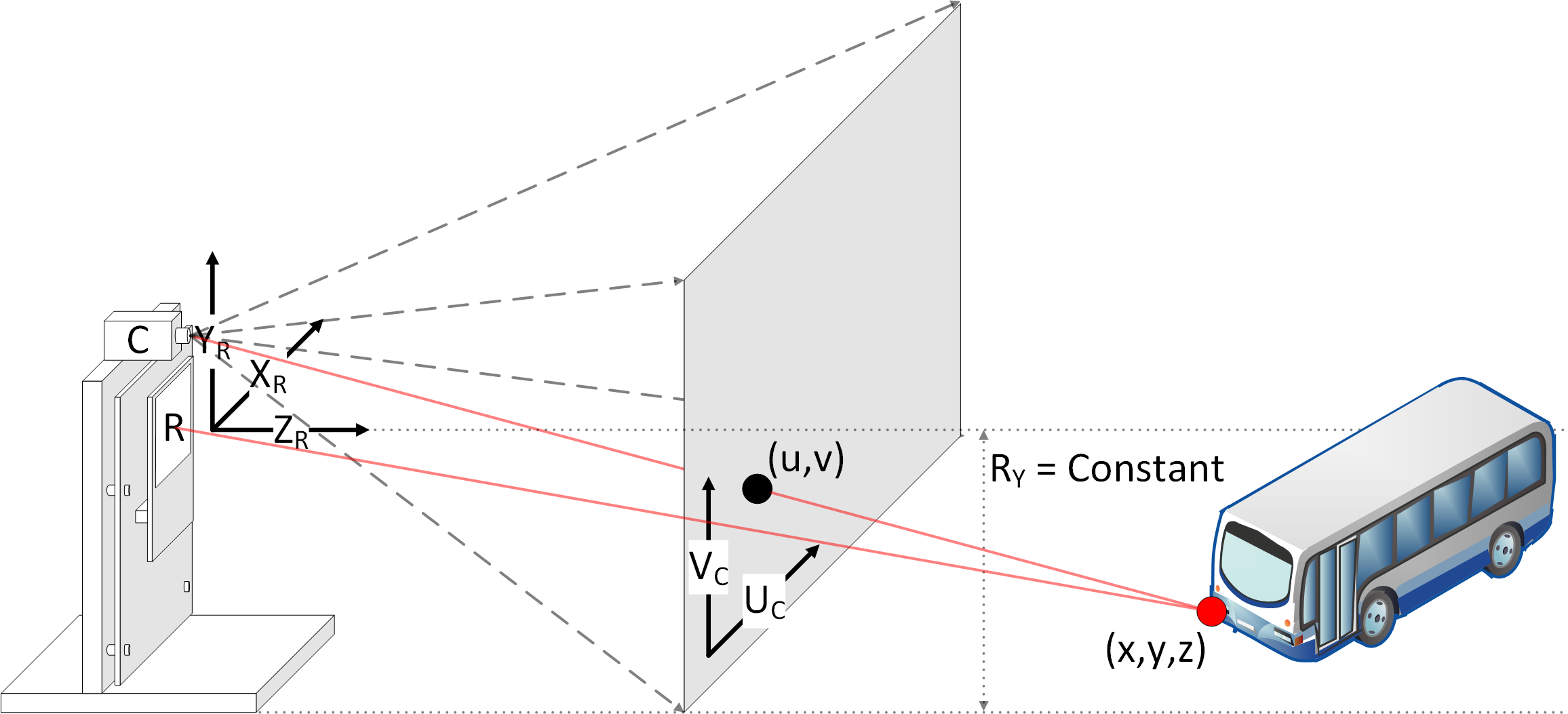}
    \caption{Evaluation Rig Calibration Diagram.}
    \label{fig:calib}
\end{figure}

From the research gathered we will leverage some of the techniques that were used to transform the two different coordinate systems to a unified field.  This will be a cornerstone to the use of this data as an input into a CNN.  Figure \ref{fig:calib} illustrates the need for calibration.  From the diagram you can see that the camera data exists in the $U_C$, $V_C$ coordinate system while the radar data exists in the $X_R$, $Y_R$, $Z_R$ coordinate system.  There is a constant RY which is the constant distance the radar sensor is above the ground.  The two data sets need to be merged together to accurately use the data for detection, tracking and identification. To do this we will follow a similar method as to \cite{sugimoto2004obstacle} where an object or radar reflector is swept through the radar beam and camera view to collect and correlate coordinate points.  Once the calibration data points are collected a transformation matrix, as seen in Eq. \ref{eq:calib}, can be used to convert to/from each coordinate system.  Elements $u$ and $v$ are the camera coordinates while $x$, $y$, and $z$ are radar coordinates. Elements $P_{11}$ through $P_{34}$ are the transformation matrix parameters that were collected as part of the calibration data collection.  The data must also be synchronized in a deterministic manner to properly correlate the data sets.

\begin{equation}
 \begin{pmatrix}
    u\\
    v\\
    1
 \end{pmatrix}
 =
  \begin{pmatrix}
    P_{11} & P_{12} & P_{13} & P_{14}\\
    P_{21} & P_{22} & P_{23} & P_{24}\\
    P_{31} & P_{32} & P_{33} & P_{34}
  \end{pmatrix}
  \begin{pmatrix}
    x\\
    y\\
    z\\
    1
 \end{pmatrix} 
 \label{eq:calib}
\end{equation}

\section{Experimental Results}
\label{sec:experimental}

In order to validate the correctness and effectiveness of the HODET scheme, two sets of experimental studies have been conducted. On the one hand, a proof-of-concept prototype was built using a JeVois camera and a TIDEP-01000 mmWave radar sensor, which has been tested in a real-world scenario; on the other hand, the processing algorithm has been further tested using the Oxford Radar RobotCar Dataset \cite{RadarRobotCarDatasetICRA2020}.

\subsection{Study on the Proof-of-Concept Prototype}

The initial test plan as follows:
\begin{itemize}
    \item Configure JeVois Image Sensor
    \item Use a pre-learned CNN for object detection
    \item Configure the mmWave Sensor
    \item Mechanically fix the two sensors to prepare for calibration
    \item Merge data sets from objects gathered from the JeVois then cascade this in conjunction with the Radar data for another CNN.
\end{itemize}

\begin{figure}[t]
    \centering
        \includegraphics[width=0.3\textwidth]{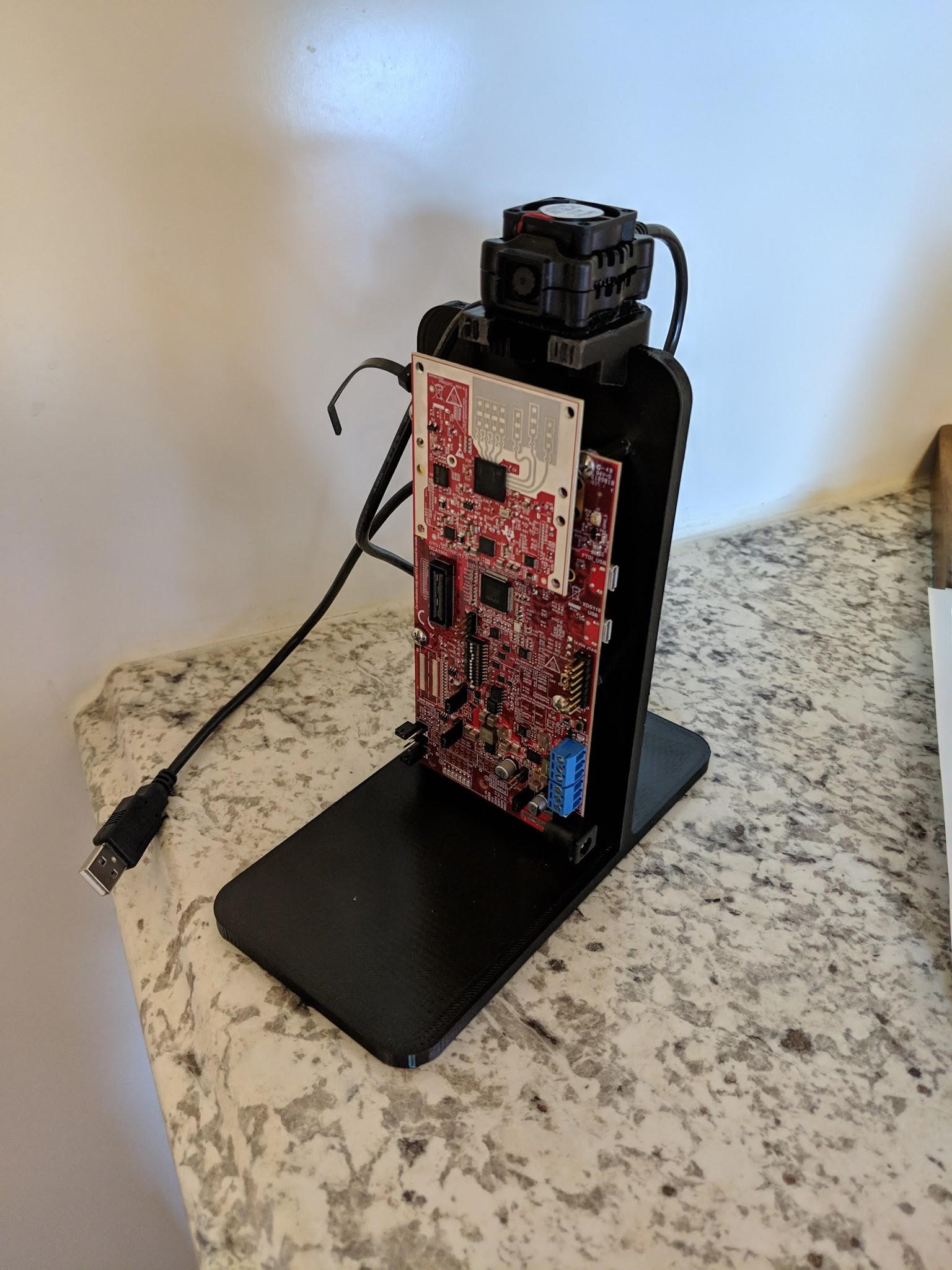}
    \caption{Evaluation Rig.}
    \label{fig:rig}
    \vspace{-10pt}
\end{figure}

The development hardware was placed on a 3D printed platform to enable the stable and consistent operation of the aperture of the CMOS imager and the radar sensor array. This is shown in Figure \ref{fig:rig}.

\begin{figure}[t]
    \centering
        \includegraphics[width=0.98\textwidth]{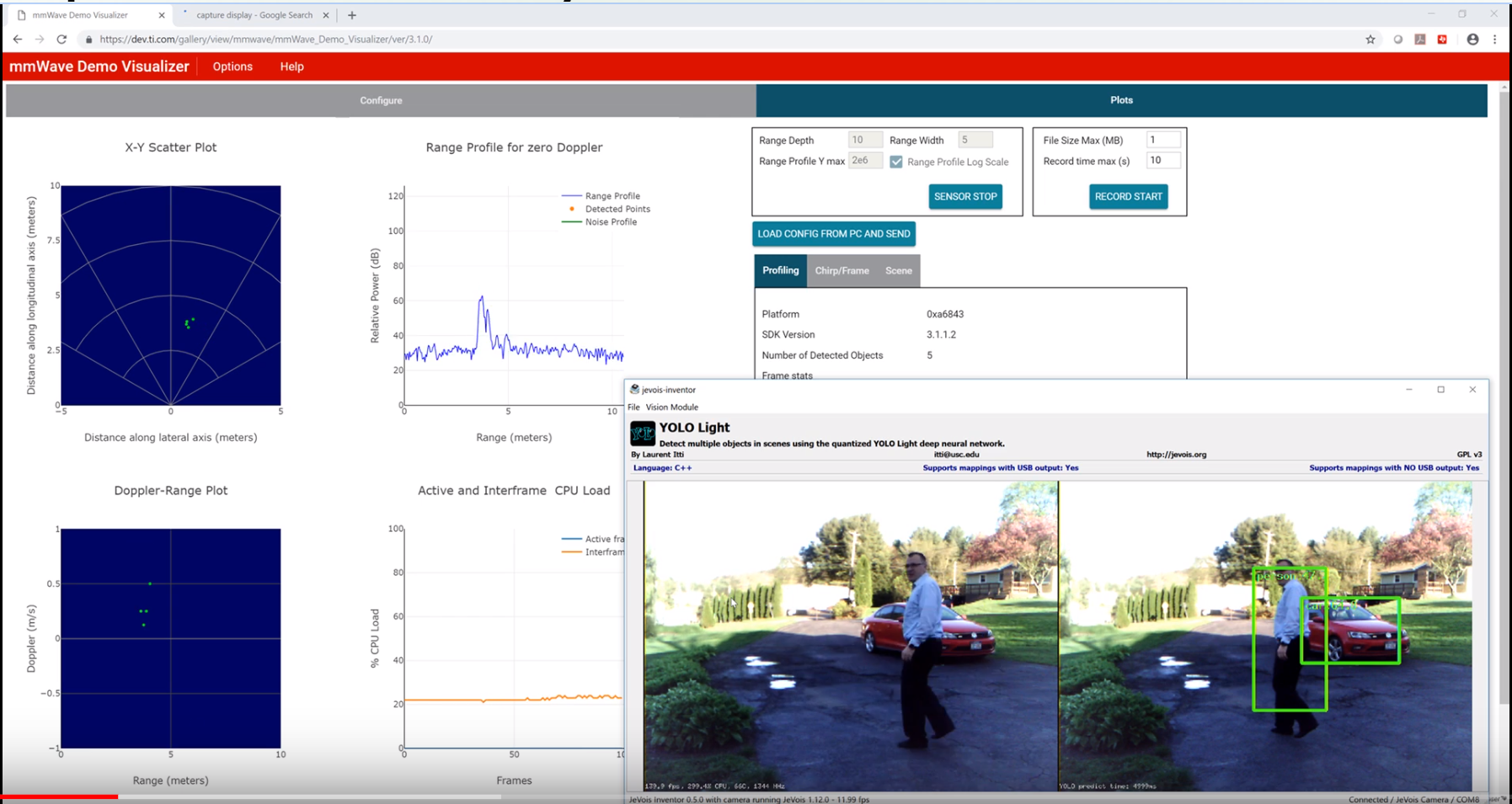}
    \caption{Initial Test Output from the Evaluation Systems.}
    \label{fig:test}
\end{figure}

The initial test of the hardware setup provides a range in the radar field of view along with the JeVois sensor running a lightweight CNN.  It is noted that the radar had a frame rate of 10 frames per second (FPS) while the inferred objects was about 2 FPS.  This is not ideal and requires some optimizations that may or may not be possible with the base hardware in the provided development kits.  Figure \ref{fig:test} shows the initial test output from the evaluation systems. Notice on the left hand side of the diagram the radar data spike denoting a detected object which can be correlated with the pedestrian on the lower right side of the diagram. Also note that the pedestrian and vehicle are detected by the JeVois camera denoted by the green boxes.

\begin{figure}[t]
    \centering
        \includegraphics[width=0.5\textwidth]{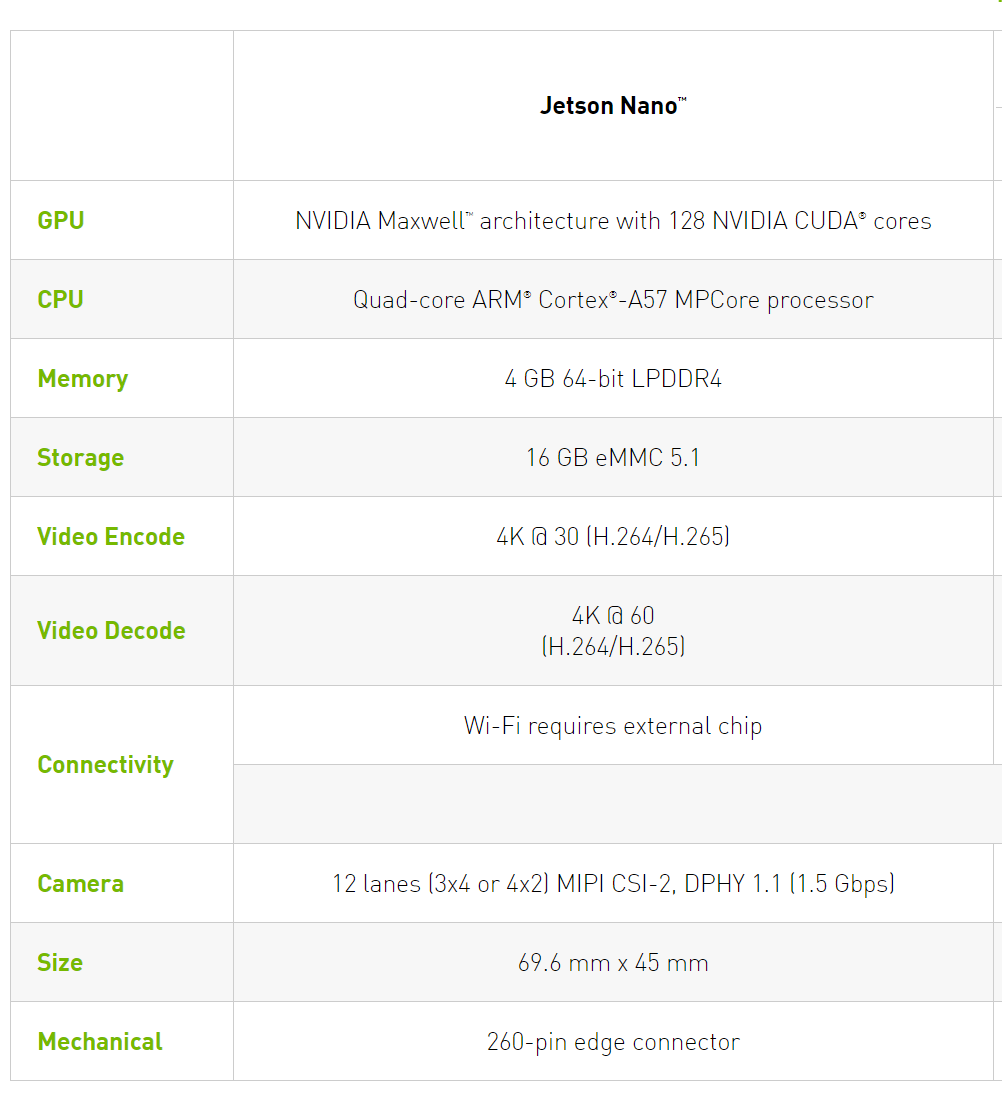}
    \caption{Jetson Nano.}
    \label{fig:jetson} 
\end{figure}

The performance of the hardware will be critical to meet our goal of a low power edge compute engine to track objects of interest with a frame rate as to not miss objects altogether. The image sensors specification are as follows \cite{jevois2020jevois}:

\begin{itemize}
    \item 1.3MP camera capable of video capture at
    \begin{itemize}
        \item SXGA (1280 x 1024): up to 15 FPS (frames/second)
        \item VGA (640 x 480): up to 30 FPS
        \item CIF (352 x 288): up to 60 FPS
        \item QVGA (320 x 240): up to 60 FPS
        \item QCIF (176 x 144): up to 120 FPS
        \item QQVGA (160 x 120): up to 60 FPS
        \item QQCIF (88 x 72): up to 120 FPS
        \item Rolling shutter, F2.8, 65\degree~horizontal field of view
    \end{itemize} 
    \item Quad-core ARM Cortex A7 processor, default clock 1.35GHz. Supports hard floating-point operations (VFPv4) and NEON SIMD multimedia instructions.
    \item Dual-core MALI-400 GPU (graphics processing unit), supports OpenGL-ES 2.0
    \item 256MB DDR3-1600 SDRAM
\end{itemize}

The above specs allow some exploration and provide a low power platform for some image sensing applications however with initial testing the camera appears to limit the frame rate to 15 FPS. This does not allow much time for the CNN which is implemented in software rather than making use of the GPU or other specialized hardware found in some new Integrated Circuits (ICs) being produced for such applications in automotive or industrial environments. Nvidia is a leading developer of such ICs that provide plenty of GPU chains to allow the parallelization of the networks. Figure \ref{fig:jetson} shows comparison hardware which will be compared to the sensor we have chosen to begin development with.

\begin{figure}[b]
    \centering
        \includegraphics[width=0.8\textwidth]{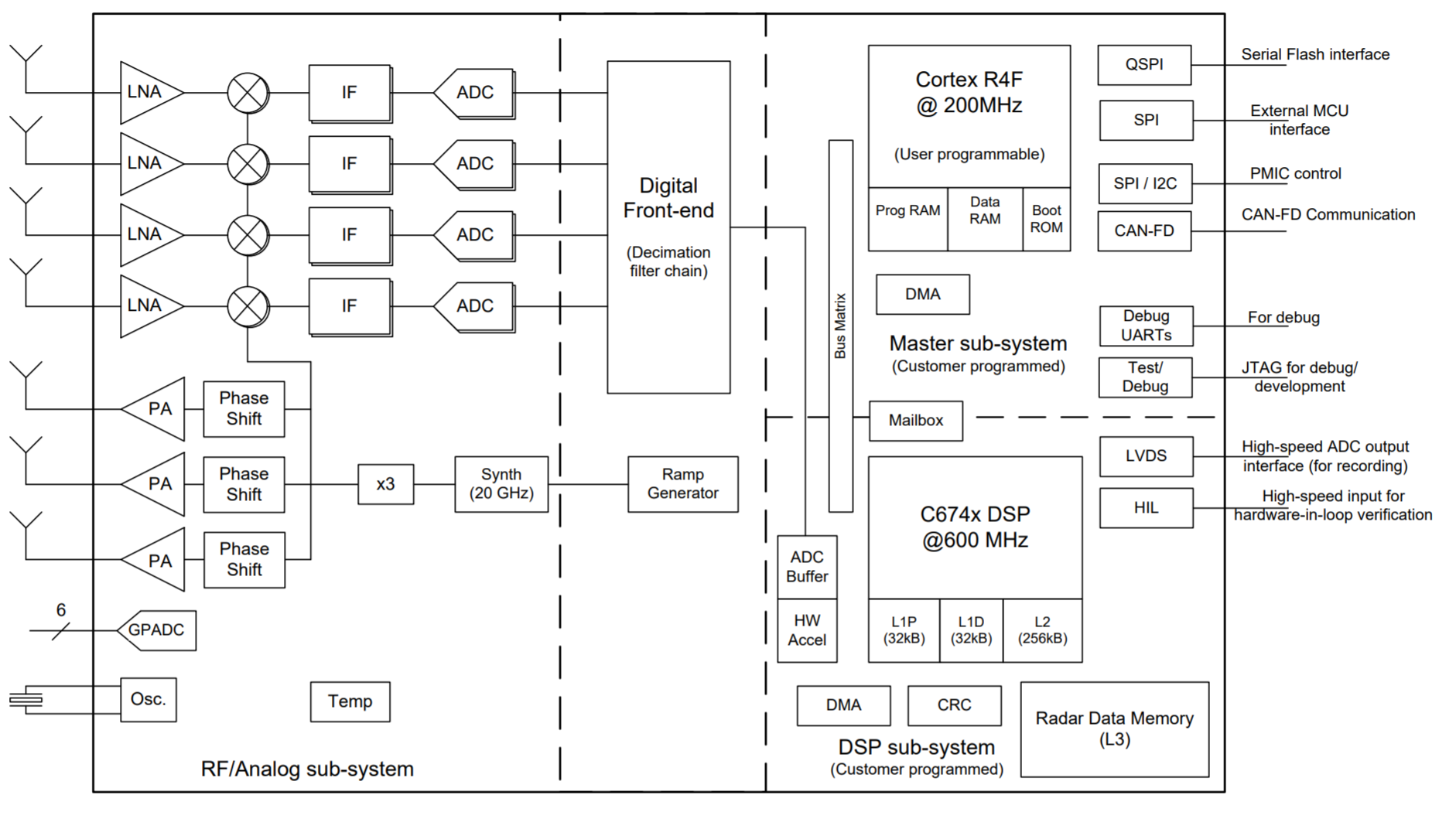}
    \caption{mmWave Sensor Block Diagram.}
    \label{fig:mmWave} 
\end{figure}

The mmWave Radar sensor appears it will meet our needs when an antenna geometry is chosen to provide the ideal FOV to match the image sensor. This initial evaluation unit provides roughly 10 meters of unambiguous range however when a custom antenna is designed the range could reach upwards of 50 to 70 meters. The IWR6843 evaluation unit provides 3 transmit (TX) and 4 receive (RX) channels to provide 108\degree~azimuth and 44\degree~of elevation. The fourth RX channel can be used to perform beam-forming techniques. Figure \ref{fig:mmWave} provides the block diagram of the mmWave sensor.

\subsection{Test using the Radar Robotcar Dataset}

\begin{figure}[t]
    \centering
        \includegraphics[width=0.8\textwidth]{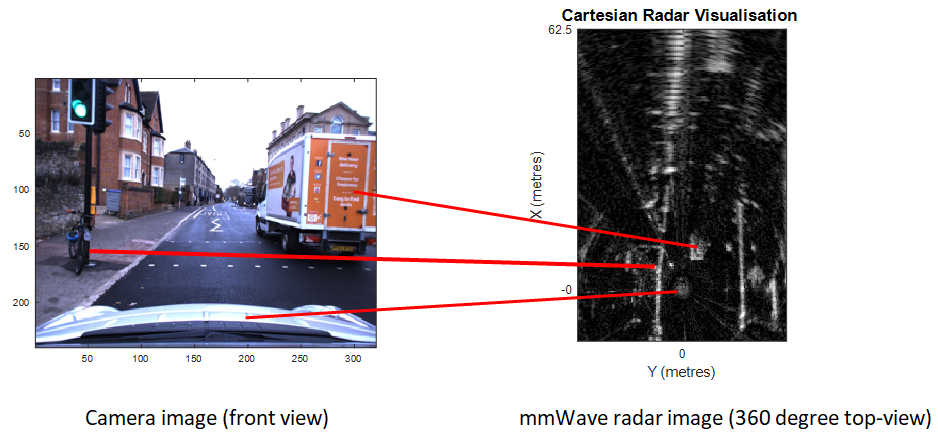}
    \caption{Sample images (a camera image and a mmWave radar image) of the Oxford Radar Robotcar dataset. Red lines indicate the same objects in these two images: truck, light pole, and the car (self). }
    \label{fig:oxford}
\end{figure}

Figure \ref{fig:oxford} shows a sample set of images (a camera image and a mmWave radar image) of the Oxford Radar Robtcar Dataset. The dataset consists of several tens hours of video and mmWave radar images, which can be used to train our proposed object detection network. Based on the trained network, we can use transfer learning to adapt the network into our own data. This provides us with a quick way of proof of concept, and can also greatly reduce our workload of acquiring training data. To furnish this, the following tasks will be conducted.

\begin{enumerate}
    \item We need to set up a calibration function to match the corresponding point positions between camera images and radar images. The original dataset does not include such a calibration example.
    
    \item We need to do a heavy hand labeling work to label typical objects in camera images and radar images for training purpose. The original dataset unfortunately does not have labeling information. 
    
    \item We need to train CNN classifiers and YOLO object detection networks from scratch because none of them had been trained directly over mmWave radar images. Retrain of the object detection YOLO network will be a nontrivial task.
\end{enumerate}

After training the object detection networks with the Oxford Radar Robotcar dataset, we will then apply transfer learning to fine tune these networks over our own data acquired by our HODET prototype. This is necessary because the mmWave radar images in the Oxford dataset are somewhat different from the our dataset. Specifically, the Oxford dataset gives $360^\circ$ top-view mmWave images because it has a $360^\circ$ revolving mmWave sensor. But it provides only azimuth and range information. In contrast, our mmWave sensor can provide zaimuth-elevation-range information but has a limited azimuth FoV with a fixed sensor.


\section{Conclusions}
\label{sec:conclusions}

There is an increasing demand for effective, efficient, and reliable surveillance solutions to maintain situational awareness (SAW) in many mission-critical delay-sensitive tasks, such as battlefield monitoring, smart public safety, disaster monitoring and recovery, etc. While the optical video surveillance system is the most popular approach, it is insufficient. 

In this paper, we propose a novel Hybrid Object DEtection and Tracking (HODET) using mmWave Radar and Visual Sensors at the edge. Through the initial demo applications provided for the JeVois camera it appears a low resolution image will need to be used to reach acceptable frame rates, this can be considered a cost of speed. Initial use of the YoloLight network has an inference time of ~500ms. The JeVois Single Board Computer does not have a direct input that could support the mmWave so this would be an improvement in a system moving forward to allow a CNN to run on one device rather than more ICs, more space and more power. Considering the small amount of data obtained using our own prototype, a larger scale data set from the Oxford Radar RobotCar Dataset has been identified, which allows us to exam the effectiveness and correctness of the core algorithms of our HODET scheme. 

This work has just begun and appears it could benefit from studying some alternate hardware systems to optimize the solution. Ultimately, a desirable hybrid sensor that meets the demands of an edge computing device can provide accurate object tracking through adverse weather with the ability to delineate objects, which could try to fool (i.e. spoof) such a camera system. There are a lot of open questions to be addressed, including the performance, security, privacy, and robustness of such a powerful but complicated system. The authors hope this preliminary study will inspire more active discussions in the community.



\bibliography{report}   
\bibliographystyle{spiejour}   

\end{document}